\input amssym.def
\font\teneufb=eufb10
\font\seveneufb=eufb7
\font\fiveeufb=eufb5
\newfam\eufbfam
\textfont\eufbfam=\teneufb
\scriptfont\eufbfam=\seveneufb
\scriptscriptfont\eufbfam=\fiveeufb
\def\bfr#1{{\fam\eufbfam\relax#1}}

\input amssym

\documentstyle[12pt]{article}
\begin{document}

\centerline{\Large\bf Smooth Loops, Generalized Coherent States  }
\vspace*{0.035truein}
\centerline{\Large\bf and Geometric Phases}
\vspace*{0.37truein}
\centerline{\bf Alexander I. Nesterov\footnote{ On leave from
Department of Theoretical Physics, Krasnoyarsk University,
Krasnoyarsk, Russian Federation.}$^,$\footnote{Departamento de F\'\i sica,
CUCEI, Universidad de Guadalajara, Corregidora, 500, S.R., Guadalajara
44840, Jalisco, M\'exico; e-mail: nesterov@udgserv.cencar.udg.mx}
and Lev V. Sabinin\footnote{On leave from Department of Mathematics,
Friendship of Nations University, Moscow, Russian Federation}$^,$\footnote
{Departamento de Matem\'aticas, Universidad Michoacana de San Nicol\'as de
Hidalgo, Morelia 58260, Michoac\'an, M\'exico; e-mail:
lev@jupiter.ccu.umich.mx}}

\vspace*{0.21truein}
\begin{abstract}
 A description of generalized coherent states and geometric
phases in the light of the general theory of smooth loops is given.

\end{abstract}

\noindent
Running head: Smooth Loops, Generalized Coherent States  and Geometric
Phases \\

\vspace*{5pt}

\section{INTRODUCTION}

Berry (1984) showed that a quantum system whose
Hamiltonian depends on some parameters $\lambda^a$ and which slowly evolves
in time in such a way that during the evolution the state of the
system traces out a closed curve $C$ in the space of these
parameters, the wave function can get an additional geometrical phase
$\gamma(C)$. This geometric phase depends on motion of the system in
the space of parameters and independent of the dynamical evolution.

	Later Aharonov and Anandan (1987) generalized Berry's result to
any cyclic evolution of the quantum system by giving up the assumption of
adiabaticity. Samuel and Bhandari (1988) introduced the geometric phase for
an arbitrary case, the evolution of the quantum system need be neither
unitary nor cyclic. These results have a simple geometric interpretation,
the evolution of the geometric phase is determined by the natural
connection in a fiber bundle over a space of rays (Simon, 1983; Aharonov
and Anandan, 1987; Samuel and Bhandari, 1988).

	Some years ago Giavarini and Onofri (1988) have described Berry's
phase using generalized coherent states. Generalized coherent states are
parametrized by points of homogeneous space where the group acts. These
appear naturally in  physical systems having dynamical symmetries, for
instance, in certain nonstationary systems such as a quantum oscillator in
a variable external field, or spin in a varying magnetic
field (Perelomov, 1972, 1986).

	In this paper we present an analysis of generalized coherent states
and geometric from the theory of smooth loops (Sabinin, 1988, 1991, 1995)
point of view.

\section{SMOOTH LOOPS AND GEOMETRIC \protect\newline PHASES}

	Let $G$ be an arbitrary Lie group and  $T(g)$ its unitary irreducible
representation acting in Gilbert space $\cal H$. Consider a fixed vector
$|\psi_0 \rangle \in \cal H$ and set of vectors (states)
$\{|\psi_g \rangle\}$, where $|\psi_g \rangle=T(g)|\psi_0 \rangle$.

{\bf Definition}. (Perelomov, 1986) A system of states  $\{|\psi_g\rangle:\,
|\psi_g \rangle=T(g)|\psi_0 \rangle\}$, where $g \in G$  and $T(g)$  is a
representation of the group $G$, acting in the Hilbert space $\cal H \;$
($|\psi_0\rangle$ is a fixed vector in this space), is called generalized
coherent-state system $\{T, |\psi_0 \rangle\}$.

	Suppose $H\subset G$ is an isotropy subgroup for the vector
$|\psi_0\rangle$, that as
$
T(h)|\psi_0 \rangle = e^{{\rm i} \alpha(h)}|\psi_0\rangle, \quad
\forall h\in H.
$
It shows that the coherent state $|\psi_g\rangle$ is determined by a point
$x=x(g)=g\cdot H$ in the left coset space $G/H$. Choosing a representative
element $g(x)$  in any equivalence class $x\in X=G/H$ one gets a set of
generalized coherent states $\{|\psi_g\rangle:\;
|\psi_g\rangle=e^{{\rm i} \alpha}|\psi_{g(x)}\rangle\}$.
From the mathematical point of view we are now considering a certain left
homogeneous space $G/H$   uniquely determining the given coherent-state
system. Actually $|\psi_g\rangle$ depends not on $g\in G$ but on the left
coset $gH\in G/H$. Choosing one single element from any coset, one
obtains in such a way a cross section $Q\subset H$, $Q\cap H$ = \{single
element\}, $Q\cap H=\{1_G\}, \; Q\cdot H =G$. Such a cross section is
called a {\it transversal} ({\it quasireductant}) of a homogeneous space
$G/H$ (see (Baer, 1940, 1942; Sabinin 1972)). Because of one-to-one
correspondence between $Q$ and $G/H$ it gives us a parametrization of $G/H$
by points of $Q$. We shall use this approach further on.

	Any transversal (quasireductant) $Q$ can be equipped in canonical way
with the structure of a left loop $\langle Q,\;\star, \;
\varepsilon\rangle$ (which means a set $Q$ with the binary multiplication
$\star$, right neutral element $\varepsilon, \; x\star \varepsilon=x$ and
unique solvability of $a\star z=b,\; z=a\backslash  b$). The construction
is as follows:  $ \forall q_1,q_2\in Q:q_1\star q_2=\pi_Q(q_1\cdot q_2), \;
\varepsilon =1_G, $ where $q_1\cdot q_2$ means a product in the group $G$,
and $\pi_Q$ the projection from $G$ onto $Q$ along left cosets,
$\{\pi_Q(q_1\cdot q_2)=Q\cap [(q_1\cdot q_2)\cdot H]\}$ (see (Sabinin,
1972) for details).

{\bf Remark.} This construction frequently gives a
two-sided loop $\langle Q,\;\star, \; \varepsilon\rangle$, which means that
$a\star x=b,\; y\star c=d$ are uniquely solvable $(x=a\backslash b, y=d/c)$
and $\varepsilon\star x=x\star\varepsilon=x$.

Generally speaking the choice of transversal $Q$ is not unique, although it
is known how the structures of loops related to different reductants are
connected (Sabinin, 1972). For a Lie group $G$ with a nondegenerate Killing
metric  on Lie subgroup $H$ a transversal (quasireductant) $Q$ can be
constructed in such a way that $h\cdot Q \cdot h^{-1}\subset Q\; (\forall
h\in H)$. Such $Q$ is called a {\it reductant} (Sabinin,1972). The
standard and unique way to construct a reductant is the following:  taking
$\bfr{g}$  and $\bfr{h}$ being Lie algebras for $G$ and $H$, respectively,
one can introduce a subspace $\bfr{m}= \{\zeta\in \bfr{g},\;
\langle\zeta,\bfr{h}\rangle=0\}$ (here $\langle\zeta,\eta\rangle$ means
Killing's inner product on $\bfr g$).  Since $\langle\zeta,\eta\rangle$ is
nondegenerate on $\bfr{h}$ we get $\bfr{m}\cap\bfr{h}=\{0\},\;\mbox{and}\;
\bfr{g = m \dotplus h}\; ({\rm direct\; sum})$. Taking $Q=\{q=\exp \xi,
\xi\in {\bfr{m}}\}$, we get a smooth reductant, at least locally.

     Thus for any $g\in G$ we have the unique decomposition
$g=\pi_Q g\cdot\pi_H g=q\cdot h$, where $q=\pi_Q g\in Q,\; h=\pi_H g\in H$.
Consequently, for the representation $T$ of $G$ we have
\begin{eqnarray}
T(g)=T(\pi_Q g\cdot\pi_H g)=T(\pi_Q g)\circ T(\pi_H g) \nonumber \\
=T(q)\circ T(h)\equiv D(q)\circ T(h), \quad (q\in Q,\; h\in H).
\end{eqnarray}
(We have used $D(q)$ instead of $T(q)$ in order to emphasize that $D(q)$ is
considered only for $q\in Q$.) Calculating, further,  $T(q_1\cdot q_2);\;
q_1,q_2\in G$, we find
\begin{eqnarray}
D(q_1)\circ D(q_2) \equiv T(q_1)\circ T(q_2) = T(q_1\cdot q_2) \nonumber \\
=T(\pi_Q (q_1\cdot q_2)\cdot \pi_H (q_1\cdot q_2)) =T((q_1\star q_2)\cdot \pi_H (q_1\cdot
q_2))  \nonumber \\
=D(q_1\star q_2)\circ T(\pi_H (q_1\cdot q_2)).  \nonumber
\end{eqnarray}

According to (Sabinin, 1972), $\pi_H (q_1\cdot q_2)$  is uniquely
determined by associator $l(q_1,q_2)=L^{-1}_{q_1\star q_2}\circ
L_{q_1}\circ L_{q_2} $, where $L_a b \stackrel{def}{=}a\star b$.  Thus
$T(\pi_H (q_1\cdot q_2))=\Lambda(q_1,q_2)$ can be regarded as an associator
of our representation. As a result, we get \begin{equation} D(q_1\star
q_2)=D(q_1)\circ D(q_2)\circ \Lambda^{-1}(q_1,q_2), \; (q_1,q_2\in Q).
\label{eq:1}
\end{equation}
Note that $D(\xi_1) \divideontimes D(\xi_2)= D(\xi_1\star \xi_2)$ where
$\divideontimes$ denotes the non-associative multiplication in the
representation.

	Let $|\psi_g\rangle=D(g)|\psi_0\rangle$ be an invariant state  with
respect to adjoint transformation: $|\psi_g\rangle =Ad_g
D(q)|\psi_g\rangle$, where $Ad_g D(q)=D(g)\circ D(q)\circ D^{-1}(g)$,
$g,q\in Q$. Using equation (\ref{eq:1}), we get
\begin{equation}
|\psi_g\rangle = D(g\star q)\circ\Lambda(g,q)|\psi_0\rangle.
\end{equation}
For the infinitesimal transformations $g+dg=R_{\delta q}g$, where
$R_{\delta q} g =g\star \delta q$ is the right action, we find (let
$f(p)$ be a smooth mapping: $\frak M \stackrel{f}{\longrightarrow}\frak
M'$.  We use $f_\ast$ for the tangent mapping: $T_p(\frak M)
\stackrel{f_\ast}{\longrightarrow}T_{f(p)}(\frak M)$)
\begin{equation}
d|\psi_g\rangle =d D(g)|\psi_0\rangle =D(g)(\Lambda(g,0))_{\ast}
(L^{-1}_g)_{\ast}d g|\psi_0\rangle.  \label{eq:2}
\end{equation}
Let $g=g(t)$ be a curve in the space $Q$. Then equation (\ref{eq:2}) yields
\begin{equation}
(d/dt)|{\psi}_g\rangle = -D(g)(\Lambda(g,0))_{\ast}D^{-1}(g)|\psi_g\rangle
(L^{-1}_g)_{\ast}dg/dt,
\label{psi}
\end{equation}
which is the differential equation for invariant state $|\psi_g\rangle$.
Multiplying by $\langle\psi_g|$ we obtain
\begin{equation}
\langle\psi_g|(d/dt)|\psi_g\rangle =
-\langle\psi_0|(\Lambda(g,0)_{\ast}|\psi_0\rangle (L^{-1}_g)_{\ast}d{g}/dt.
\label{eq:3}
\end{equation}

	One can write (\ref{psi}) as
\begin{equation}
(d/dt)|{\psi}_g\rangle - {\rm i}A(t)|\psi_g\rangle = 0,
\end{equation}
where $A= {\rm i}D(g)(\Lambda(g,0))_{\ast}D^{-1}(g) (L^{-1}_g)_{\ast}dg/dt
$ is introduced.  If one makes a gauge transformation
\[
|{\psi}'_g\rangle=
e^{{\rm i} \alpha(t)} |{\psi}_g\rangle ,
\]
then the $A$ field transforms as
\[
A'= A + d\alpha/dt,
\]
i.e. as proper gauge potential. So equation (\ref{psi}) gives a definition 
of a parallel transport in the Hilbert space expressed in terms of the 
invariant states with respect to adjoint action of the loop $Q$.

	Suppose that the normalized state $|\psi(t)\rangle=e^{{\rm i}
\varphi(t)}|\psi_{g(t)}\rangle$ evolves according to the Schr\"{o}dinger
equation ${\rm i}(d/dt)|{\psi}(t)\rangle=\hat{H}|\psi(t)\rangle$; hence
$d{\varphi}/dt =- \langle\psi_g|\hat H|\psi_g\rangle +{\rm i}
\langle\psi_g|(d/dt)|{\psi}_g\rangle$. For cyclic evolution of a quantum
system,  $|\psi_{g(\tau)}\rangle=|\psi_{g(0)}\rangle$, $g(\tau)=g(0)$, the
total phase being $\varphi=\gamma - \delta$, where \[ \delta=- \int^\tau_0
\langle\psi_g|\hat H|\psi_g\rangle d t \] is the dynamical phase and \[
\gamma={\rm i}\oint_C\langle\psi_g|d|{\psi}_g\rangle
\]
is the Aharonov--Anandan (AA) geometric phase (Aharonov and Anandan,
1987).  Using equation(\ref{eq:3}), we get
\begin{equation}
\gamma = - {\rm i}\oint_C\langle\psi_0|(\Lambda(g,0))_{\ast}|\psi_0\rangle
(L^{-1}_g)_{\ast} d g.
\label{eq:4}
\end{equation}
Hence $\gamma$ defined by equation (\ref{eq:4}), is independent of $\varphi$ and
uniquely determined by the associator and the curve $C\in Q$.
Thus we come to conclusion that the A-A geometric phase is originated
by nonassociativity.

\section{Examples}

Now we consider two applications of the above theory to the generalized
coherent states for the groups $SU(1,1)$ and $SU(2)$.

{\it 1. Group $SU(1,1)$}. The group $SU(1,1)$ can be consider as a group of
transformations of the complex plane $\Bbb C$.  The action of this group is
intransitive and the complex plane $\Bbb C$ is divided into three orbits:
${\Bbb C}_+ =\{z:|z|<1\}, \; {\Bbb C}_- =\{z:|z|>1\}, \; {\Bbb C}_0
=\{z:|z|=1\}$.  The Lie algebra corresponding to the group $SU(1,1)$ is
spanned by generators $K_0,K_{\pm}$ with the commutation relations:
$[K_0,K_{\pm}]=K_{\pm}, \;[K_-,K_+]=2K_0$. Let us now restrict ourselves to
consideration of ${\Bbb C}_+$. The set of complex numbers $\xi,\eta\in
{\Bbb C}_+$ with the operation $\star$  forms a two-parametric loop
$QH(2)$ (Nesterov and Stepanenko, 1986; Nesterov, 1989, 1990)
\begin{equation}
L_\zeta\eta\equiv\zeta\star\eta=\frac{\zeta+\eta}{1+\overline\zeta\eta},
\label{eq:7}
\end{equation}
here $\overline\zeta$ is the complex conjugate number $(\zeta=x+{\rm i} y,
\quad \overline\zeta=x-{\rm i} y)$. The associator
$l(\zeta,\eta)=L^{-1}_{\zeta\star\eta}\circ L_\zeta\circ
L_\eta$ on $QH(2)$ is determined by
\begin{equation}
l(\zeta,\eta)=\frac{1+\zeta\overline\eta}{1+\eta\overline\zeta}
\end{equation}
and  can be written also as $l(\zeta,\eta)=\exp({\rm i}  \alpha), \quad
\alpha =2\arg(1+\zeta\overline\eta)$. This loop is isomorphic to the
geodesic loop of a two-dimensional Lobachevski space realized as the upper
part of two-sheeted unit hyperboloid $H^2$ (Sabinin, 1991, 1995). The
isomorphism is established by exponential mapping $\zeta=e^{{\rm
i}\varphi}\tanh\frac{\tau}{2}$ where $(\tau,\varphi)$ are inner
coordinates  on $H^2$.

	The group $SU(1,1)$ is non-compact and all its unitary irreducuble
representations are infinite-dimensional. We shall consider only a discrete
representation, which is determined by a single number
$k=1,3/2,2,5/2,\dots$  and $K_0|k,\mu|\rangle = \mu|k,\mu|\rangle
,\;\mu=k+m$, where $m$ is an integer ($m\geq 0$).
The operators $D(\xi)$, determined as
\begin{equation}
D(\xi) =\exp(\xi K_+ - \overline\xi K_-), \quad
\xi=-\frac{\tau}{2} e^{-{\rm i}\varphi},
\end{equation}
form a nonassociative representation of the loop $QH(2)$ with the
multiplication law (see (2))
\begin{eqnarray}
D(\xi_1) \divideontimes D(\xi_2)=
D(\xi_1)\circ D(\xi_2)\circ \exp(-{\rm i} {\alpha}K_0)
\label{eq:6}
\end{eqnarray}
where $\alpha =- {\rm i}\ln(l(\zeta_1,\zeta_2))$ and
$\zeta_1=e^{{\rm i}\varphi_1}\tanh|\xi_1|, \;
\zeta_2=e^{{\rm i}\varphi_2}\tanh|\xi_2|$.
The canonical set $\{|\zeta\rangle\}$ of coherent
states,  corresponding to the choice $|\psi_0\rangle = |k,k\rangle$,
is (Perelomov, 1986)
\begin{equation}
|\zeta\rangle = (1-|\zeta|^2)^k \exp(\zeta K_+)|\psi_0\rangle,
\label{eq:5}
\end{equation}
where $\zeta=e^{{\rm i}\varphi}\tanh\frac{\tau}{2}$. The infinitesimal
operators in this representation are
\[
\langle\zeta|K_0|\zeta\rangle =k\frac{1+|\zeta|^2}{1-|\zeta|^2},\quad
\langle\zeta| K_+|\zeta\rangle =2k\frac{\overline\zeta}{1-|\zeta|^2}.
\]

	Now let us compute the geometric phase $\gamma$ using (\ref{eq:4})
and (\ref{eq:6}). We find
\begin{equation}
\gamma(C) = \langle\psi_0|K_0|\psi\rangle\oint_C \delta\alpha
\end{equation}
where $\delta\alpha={\rm i}\ln((1+\overline\zeta
\delta\zeta)/(1+\zeta \delta\overline\zeta))={\rm i} (\overline\zeta
\delta\zeta - \zeta \delta\overline\zeta)$. Applying (\ref{eq:7}), we
obtain $\delta\zeta=d\zeta/(1-|\zeta|^2)$. This yields
\begin{equation}
\gamma(C) =\langle\psi_0|K_0|\psi_0\rangle\oint_C \delta\alpha ={\rm i}k
\oint_C\frac{\overline\zeta d\zeta - \zeta d\overline\zeta} {1-|\zeta|^2}=
-k A,
\label{eq:12}
\end{equation}
where $A$ is the area of the hyperboloid's surface corresponding to
the region bounded by the closed path $C\in{\Bbb C}_+$, with taking
into account the equation $K_0|\psi_0\rangle =k|\psi_0\rangle$.

On the basis of  equation (\ref{eq:5}) one can calculate $\gamma$ directly:
\begin{eqnarray}
\frac{d\gamma(C)}{dt}={\rm i}\langle\zeta|(d/dt)|\zeta\rangle
= {\rm i}k \frac{\overline\zeta d\zeta/dt-\zeta {d\overline\zeta}/dt}{1-
|\zeta|^2}  =-2k\frac{d\varphi}{dt}\sinh^2{\frac{\tau}{2}}
\end{eqnarray}
and the total phase is the same as in (\ref{eq:12}),
\begin{equation}
{\gamma}(C)={\rm i}\oint_C\langle\zeta|d\zeta\rangle =
-2k\oint_C\sinh^2\frac{\tau}{2}\; d\varphi=-k A.
\end{equation}
	The unit hyperboloid $H^2$ can be considered as as the phase manifold
of the quantum parametrically excited oscillator (Perelomov, 1986). Hence
(\ref{eq:12}) determines the AA geometric phase for cyclic evolution of
this oscillator.

{\it 2. Group $SU(2)$}. The consideration is similar to that used for
$SU(1,1)$. The essential difference between the group $SU(2)$ and $SU(1,1)$ 
is that the first is compact and simply connected, while the second is 
neither. The Lie algebra of $SU(2)$ is spanned by generators $J_0,J_{\pm}$ 
with the standard commutation relations:  $[J_0,J_{\pm}]=J_{\pm}, 
\;[J_-,J_+]=-2J_0$.  Any unitary irreducible representation of the group 
$SU(2)$ is determined by nonnegative integer or half-integer $j$. In the 
space of representation ${\cal H}^j$ we shall use the canonical basis 
$|j,\mu\rangle \; -j\leq \mu\geq j$ of eigenvectors of the operator $J_0:\; 
J_0 |j,\mu\rangle = \mu|j,\mu\rangle$. The generalized coherent states 
correspond to points of the two-demensional sphere $S^2$ and the set of 
operators \{$D(\xi)$\} is given by (Perelomov, 1986) 
\begin{equation} 
D(\xi) =\exp(\xi J_+ - \overline\xi
J_-), \quad \xi= -\frac{\theta}{2} e^{-{\rm i}\beta}
\end{equation}
where $\beta=\pi -\varphi$ and $\theta,\varphi$ are the usual spherical
coordinates. This corresponds to the stereographic projection onto the
complex plane $\Bbb C$ from the north pole of the sphere. Applying the
operator $D(\xi)$ in its normal form,
\begin{equation}
D(\xi) =\exp(\zeta J_+)\exp(\eta J_0)\exp(- \overline\zeta J_-),
\end{equation} where
$\zeta=e^{{\rm i}\varphi}\tan\frac{\theta}{2}, \quad \eta
=\ln(1+|\zeta|^2)$, to the state vector $|\psi_0\rangle = |j,-j\rangle$,
one gets the set of coherent states (Perelomov, 1986)
\begin{eqnarray}
|\zeta\rangle = (1+|\zeta|^2)^{-j} \exp(\zeta J_+)|j,-j\rangle,
\nonumber \\
\langle\zeta|J_0|\zeta\rangle =-j\frac{1-|\zeta|^2}{1+|\zeta|^2},\quad
\langle\zeta| J_+|\zeta\rangle =2j\frac{\overline\zeta}{1+|\zeta|^2}.
\label{eq:8}
\end{eqnarray}

	The sphere $S^2$ admits a natural quasigroup structure, namely, $S^2$
is a local two-parametric loop $QS(2)$ (Nesterov and Stepanenko,
1986; Nesterov, 1989, 1990)
\begin{equation}
L_\zeta\eta\equiv\zeta\star\eta=\frac{\zeta+\eta}{1-\overline\zeta\eta}
\label{eq:9}
\end{equation}
where $\zeta,\eta\in {\Bbb C}$  and the isomorphism between points of the
sphere and the complex plane ${\Bbb C}$ is established by the stereographic
projection from the north pole of the sphere: $\zeta = e^{{\rm i}\varphi}
\tan \frac{\theta}{2}$.  The associator is determined by
\begin{equation}
l(\zeta,\eta)=\frac{1-\zeta\overline\eta}{1-\eta\overline\zeta}
\end{equation}
and  can be written also as $l(\zeta,\eta)=\exp({\rm i}\alpha), \quad
\alpha =2\arg(1-\zeta\overline\eta)$. The operators $D(\xi)$ form a
nonassociative representation of $QH(2)$:
\begin{eqnarray}
D(\xi_1)\divideontimes D(\xi_2)= D(\xi_1)\circ D(\xi_2)\circ \exp(-{\rm i}
{\alpha}K_0)
\label{eq:10}
\end{eqnarray}
where $\alpha =- {\rm i}\ln(l(\zeta_1,\zeta_2))$ and
$\zeta_1=e^{{\rm i}\varphi_1}\tan|\xi_1|,\;
\zeta_2=e^{{\rm i}\varphi_2}\tan|\xi_2|$.

	Let us compute $\gamma$ using (\ref{eq:4}). Applying (\ref{eq:10}),
we find
\begin{equation}
\gamma(C)=\langle\psi_0|J_0|\psi_0\rangle\oint_C
\delta\alpha
\end{equation}
where $\delta\alpha={\rm i}\ln((1-\overline\zeta \delta\zeta)/(1-\zeta
\delta\overline\zeta))={\rm i} (\overline\zeta \delta\zeta - \zeta
\delta\overline\zeta)$. From (\ref{eq:9}) we get
$\delta\zeta=d\zeta/(1+|\zeta|^2)$. Now taking into account
$J_0|\zeta\rangle=-j|\zeta\rangle$, we obtain
\begin{equation}
\gamma(C) =\langle\psi_0|J_0|\psi_0\rangle\oint_C \delta\alpha ={\rm i}j
\oint_C\frac{\overline\zeta d\zeta - \zeta d\overline\zeta} {1+|\zeta|^2}=
-j\Omega,
\label{eq:11}
\end{equation}
where $\Omega$ is the solid angle corresponding to the contour $C$.
Using (\ref{eq:8}) and the definition
\[
\gamma = {\rm i}\oint_C \langle\zeta|d|\zeta\rangle,
\]
one can calculate $\gamma$ directly. This yields the same result
$\gamma=-j\Omega$.

	Actually, the sphere $S^2$ can be considered as the phase manifold of
the spin (Perelomov, 1986) and, consequently, equation (\ref{eq:11}) gives 
the AA geometric phase for the cyclic evolution of the spin. For instance, 
the spin precession in a variable magnetic field ${\bf H}(t)$ is described 
by the Hamiltonian $\hat H = -\kappa {\bf H}(t) {\bf J}(t)$ and the 
geometric phase $\gamma=-j\Omega$ (Berry, 1984).

\section{CONCLUDING REMARKS}

	The above discussion shows that the generalized coherent states
actually are detemined by points of the corresponding smooth loop $Q$ and
AA geometric phase is originated by nonassociativity of the operation
(multiplication) in $Q$. Our approach can be applied also to evolution of a
quantum system, that neither unitary nor cyclic. In this general case
when the state vector does not return to the initial ray, the method of
comparing the states is given by Pancharatnam (1975). Let $|\psi_1\rangle$
and $|\psi_2\rangle$ be any two states which are not mutually orthogonal.
The Pancharatnam phase $\beta$ is defined as
\begin{equation}
e^{{\rm i}\beta} =\frac{\langle\psi_1|\psi_2\rangle}
{\|\langle\psi_1|\psi_2\rangle\|}.
\end{equation}
Now define
\[
|\psi(\tau)\rangle =e^{{\rm i}\gamma(\tau)}|\psi(0)\rangle
=D(g(\tau))|\psi_0\rangle
\]
such that
$\gamma(1)=\beta,\; |\psi(0)\rangle=|\psi_1\rangle,\;
|\psi(1)\rangle=|\psi_2\rangle$. Then $\beta$ is given by a line integral
\begin{eqnarray}
\beta=-{\rm i}\int\langle\psi|d|\psi\rangle
= -{\rm i}\int_{g_0}^{g_1}\langle\psi_0|(\Lambda(g,0))_\ast
|\psi_0\rangle (L^{-1}_g)_\ast d g \nonumber \\
= -{\rm i}\int_{g_0}^{g_1}\langle\psi_1|D(g_1)(\Lambda(g,0))_\ast
D^{-1}(g_1)|\psi_1\rangle (L^{-1}_g)_\ast d g
\end{eqnarray}
where we set $g_1=g(1),\;g_2=g(2).$

\section*{ACKNOWLEDGEMENTS}
\noindent
We are greatly indebted to N. V. Mitskievich for reading the manuscript.
This work was supported in part by CONACYT, Grant No. 1626P-E.

\section *{References}
\noindent
\begin{description}

\item  Aharonov Y. and Anandan J. (1987). {\it Physical Review Letters},
{\bf 58}, 1593.

\item  Baer R. (1940). {\it Transactions of the American Mathematical
Society}, {\bf 46}, 110.

\item  Baer R. (1941). {\it Transactions of the American Mathematical
Society}, {\bf 47}, 435.

\item Berry M. V. (1984). {\it Proceedings of the Royal Society A},
{\bf 392}, 45.

\item Giavarini G. and Onofri E. (1988). Generalized Coherent States
and Berry's Phase, UTF Preprint.

\item Nesterov A. I. (1989).{\it Methods of Nonassociative Algebra in
Physics,} Dr. Sci Theses, Institute of Physics, Estonian Academy of
Science, Tartu.

\item Nesterov A. I. (1990). In {\it Quasigroups and Nonassociative
Algebras in Physics}, Transactions of the Institute of Physics of the
Estonian Academy of Science, {\bf 66}, 107.

\item Nesterov A. I. and Stepanenko V. A. (1986). On Methods of
Nonassociative Algebra in Geometry and Physics, L. V. Kirensky Institute
of Physics, Preprint 400F.

\item  Pancharatnam S. (1975). {\it Proceedings of the Indian Academy of
Sciience A}, {\bf 44}, 247.

\item Perelomov A. M. (1972). {\it Communications in Mathematical Physics},
{\bf 26}, 22.

\item Perelomov A. M. (1986). {\it Generalized Coherent States
and Their Applications}, Springer-Verlag, Berlin Heidelberg, New York.

\item Sabinin L. V (1970). {\it Soviet Mathematical Doklady}, {\bf 13}, 970.

\item Sabinin L. V. (1972). {\it Mathematical Notes},{\bf 12}, 799.

\item Sabinin L. V. (1988). In {\it Proceedings of Seminar on Vector and
Tensor Analysis}, {\bf 23}, 133, Moscow Univ, Moscow.

\item Sabinin L. V. (1991). {\it  Analytic Quasigroups and Geometry},
Friendship of Nations University, Moscow.

\item Sabinin L. V. (1995). {\it Russian Mathematical Survey}, {\bf 49},
172.

\item  Samuel J. and Bhandari R. (1988). {\it Physical Review Letters},
{\bf 60}, 2339.

\item  Simon B. (1983). {\it Physical Review Letters}, {\bf 51}, 2167.

\end{description}

\end{document}